# Evidence for Superlattice Dirac Points and Space-dependent Fermi Velocity in Corrugated Graphene Monolayer

Hui Yan[1,§], Zhao-Dong Chu[1,§], Wei Yan[1], Mengxi Liu[2], Lan Meng[1], Mudan Yang[1], Yide Fan[1], Jiang Wang[1], Rui-Fen Dou[1], Yanfeng Zhang[2,3], Zhongfan Liu[2], Jia-Cai Nie[1], and Lin He[1,*]

[1] Department of Physics, Beijing Normal University, Beijing, 100875, People's Republic of China
[2] Center for Nanochemistry (CNC), College of Chemistry and Molecular Engineering, Peking University, Beijing 100871, People's Republic of China
[3] Department of Materials Science and Engineering, College of Engineering, Peking University, Beijing 100871, People's Republic of China

Recent studies show that periodic potentials can generate superlattice Dirac points at energies $\pm\hbar v_F|\mathbf{G}|/2$ in graphene ($v_F$ is the Fermi velocity of graphene and $\mathbf{G}$ is the reciprocal superlattice vector). Here, we perform scanning tunneling microscopy and spectroscopy studies of a corrugated graphene monolayer on Rh foil. We show that the quasi-periodic ripples of nanometer wavelength in the corrugated graphene give rise to weak one-dimensional (1D) electronic potentials and thereby lead to the emergence of the superlattice Dirac points. The position of the superlattice Dirac point is space-dependent and shows a wide distribution of values. We demonstrated that the space-dependent superlattice Dirac points is closely related to the space-dependent Fermi velocity, which may arise from the effect of the local strain and the strong electron-electron interaction in the corrugated graphene.

Since the laboratory realization of graphene in 2004 [1], this two-dimensional honeycomb lattice of carbon atoms has motivated intense theoretical and experimental investigations of its properties [2-8]. It was demonstrated that the electronic chirality (the spinorlike structure of the wavefunction) is of central importance to many of graphene's unique electronic properties [3,9-12]. Recently, a number of theoretical studies predicted that the chiral nature of charge carriers results in highly anisotropic behaviours of massless Dirac fermions in graphene under periodic potentials and generates new Dirac points at energies $E_{SD} = \pm\hbar v_F|\mathbf{G}|/2$ in graphene superlattice (here $v_F$ is the Fermi velocity of graphene and $\mathbf{G}$ is the reciprocal superlattice vector) [13-16]. Despite these suggestive findings [13-16] and many other interesting physics [17-22] in graphene superlattice, the experimental study of this system is scarce due to the difficulty in fabricating graphene under nano-scale periodic potentials [23]. Until recently, it was demonstrated that graphene superlattice (corrugated graphene or moiré pattern) induced between the top graphene layer and the substrate (or the underlayer graphene) acts as a weak periodic potential, which generates superlattice Dirac points at an energy determined by the period of the potential [24-26]. These seminal experiments provide a facile method to realize graphene superlattice and open opportunities for superlattice engineering of electronic properties in graphene.

In this Letter, we address the electronic structures of a corrugated graphene monolayer on Rh foil. We show that the quasi-periodic ripples of nanometer wavelength give rise to a weak one-dimensional (1D) electronic potential in graphene. This 1D potential leads to the emergence of the superlattice Dirac points $E_{SD}$, which are manifested by two dips in the density of states, symmetrically placed at energies flanking the pristine graphene Dirac point $E_D$. The position of $E_{SD}$ is space-dependent and shows a wide distribution of values. Our experimental result demonstrates that the space-dependent $E_{SD}$ is closely related to the space-dependent Fermi velocity, which is attributed to the effect of the local strain and the strong electron-electron interaction in the corrugated graphene.

The graphene monolayer was grown on a 25 micron thin polycrystalline Rh foil, which is mainly (111) oriented, via a traditional ambient pressure chemical vapor deposition (CVD) method, as reported in a previous paper [27]. The sample was synthesized at 1000 ºC and the growth time was varied from 3 to 15 min for controlling the thickness of graphene. The thickness of the as-grown graphene was further characterized by Raman spectra measurements [27] and in this paper we focus on the structure and electronic properties of graphene monolayer. The as-grown sample was cooled down to room temperature and then transferred into the ultrahigh vacuum condition for further scanning tunneling microscopy (STM) characterizations. The STM system was an ultrahigh vacuum four-probe scanning probe microscope from UNISOKU. All STM and scanning tunneling spectroscopy (STS) measurements were performed at liquid-nitrogen temperature and the images were taken in a constant-current scanning mode. The STM tips were obtained by chemical etching from a wire of Pt(80%) Ir(20%) alloys. Lateral dimensions observed in the STM images were calibrated using a standard graphene lattice. The STS spectrum, i.e., the dI/dV-V curve, was carried out with a standard lock-in technique using a 957 Hz alternating current modulation of the bias voltage.

Due to thermal expansion mismatch between graphene and the substrate (the Rh foil contracts, whereas the graphene layer expands during the cooling process), defect-like wrinkles and ripples tend to evolve along the boundaries of crystalline terraces for strain relief [28,29]. Very recently, this thermal strain engineering was used to generate (sub)nanometer-wavelength periodic ripples in

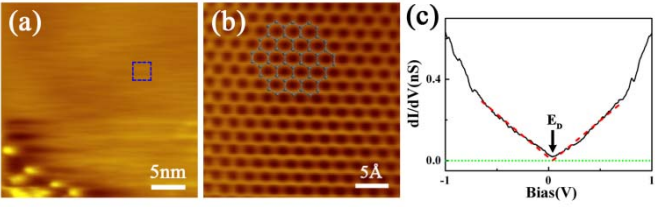

FIG. 1 (color online). (a) A STM image of a flat graphene monolayer on Rh foil ($V_{sample}$ = 550 mV and I = 6.26 pA). The bright dots around the bottom left corner arise from charge accumulation at the boundary of graphene. (b) Atomic-resolution image of graphene in the blue frame of panel (a) ($V_{sample}$ = 600 mV and I = 14.8 pA). The atomic structure of graphene is overlaid onto the STM image. (c) A typical tunneling spectrum, i.e., dI/dV-V curve, recorded on the graphene monolayer. The Dirac point, as marked by the black arrow, is located at $E_D$ ~ 38 mV indicating a slight charge transfer between the graphene and substrate. Around the Dirac point, the density of states increases linearly with the energy, which is identical to that in pristine graphene. The dashed and dotted lines are the guide to eyes.

graphene [30] and it was demonstrated that the strained graphene structures modify the local electronic structures dramatically [30-34]. Theoretically, it was predicted that corrugated graphene could lead to an electronic superlattice with a period set by the corrugation wavelength [35]. Motivated by this proposal, we address the electronic structures in graphene with quasi-periodic ripples of nanometer wavelength by STM and STS.

Figure 1(a) shows a STM image of a flat area of graphene monolayer on polycrystalline Rh foil. No periodic moiré superstructures can be seen and almost identical feature has been observed in several tens flat areas of graphene monolayer on Rh foil. This feature differs quite from that of both the graphene bilayer on Rh foil and the graphene monolayer on a (111) surface of single-crystal Rh [27]. For the graphene bilayer, the misorientation between the bilayer usually results in moiré superstructures with different periods [27]. For the latter case, the strong C-Rh covalent bond and the lattice mismatch between graphene (0.246 nm) and Rh(111) (0.269 nm) could lead to hexagonal moiré superstructures with the periodicity of 2.9 nm [27,36-38]. The absence of moiré superstructures, as shown in Fig. 1(a), indicates that the coupling between graphene and the Rh foil is much weaker than that of monolayer graphene on a single-crystal Rh [27]. Fig. 1(b) shows an atomic resolution STM image of the graphene, where a clear honeycomb lattice is observed. Fig. 1(c) shows a typical STS spectrum of the sample. The tunnelling spectrum gives direct access to the local density of states (LDOS) of the surface at the position of the STM tip. The linear DOS around the Dirac point $E_D$ consists well with that of the pristine graphene. The position of the Dirac point $E_D$ is slightly above the Fermi level, suggesting charge transfer between the graphene and the substrate [39-41].

As mentioned above, a corrugated graphene monolayer with quasi-periodic ripples of nanometer wavelength is easy to be found along the boundaries of crystalline terraces of Rh foil. Fig. 2(a) shows a typical corrugated graphene with quasi-periodic ripples (see Fig. S1 in the

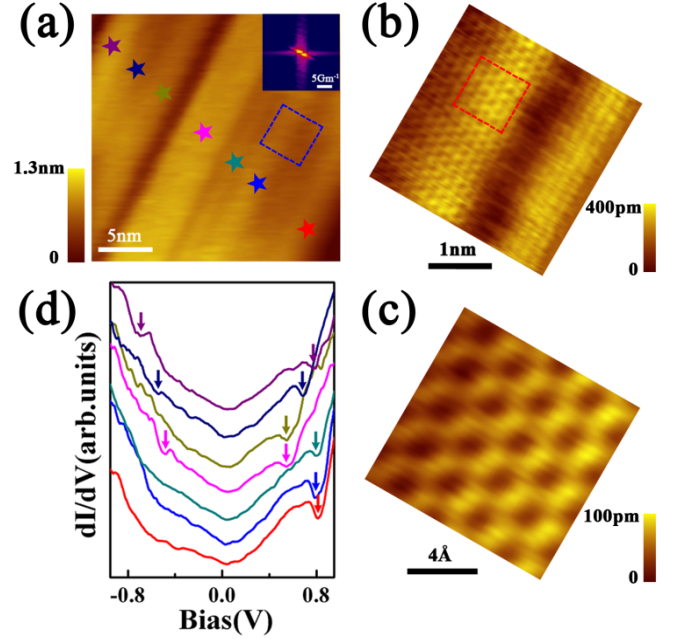

FIG. 2 (color online). (a) A STM topography of corrugated graphene with quasi-periodic ripples of nanometer wavelength on Rh foil ($V_{sample}$ = 440 mV and I = 10.31 pA). The inset is Fourier transform of the main panel showing the reciprocal-lattice of the quasi-periodic ripples. The scale bar of the inset is 5 $Gm^{-1}$. (b) Zoom-in topography of the blue frame in (a) shows a 1D superlattice ($V_{sample}$ = 393 mV and I = 10.31 pA). (c) Atomic-resolution image of the red frame in panel (b) shows a honeycomb lattice ($V_{sample}$ = 393 mV and I = 10.31 pA). (d) Tunneling spectra, i.e., dI/dV-V curves, recorded at different positions, as marked by the stars with different colors, in panel (a). The spectra were vertically offset for clarity. The positions of the superlattice Dirac points, which contribute to the dips of the local density of states, are indicated by the arrows.

supplemental material [42] for a line profile of the quasi-periodic ripples). The average width of these ripples is estimated to be about 3.5 nm and the height of these ripples is usually smaller than 1 nm. The inset is the Fourier transform showing the reciprocal-lattice of the ripples. Figure 2(b) and (c) show atomic-resolution images of the ripples and only a honeycomb lattice is observed, suggesting that the local curvature of the ripples does not break the six-fold symmetry of the graphene lattice. This result also implies that there is no gap opening at the Dirac point. In literature, a triangular lattice was observed on graphene wrinkle ~10 nm in width and ~ 3 nm in height [43]. The triangular lattice along the wrinkle may arise from its large local curvature (strain) that breaks the six-fold symmetry of the lattice.

Figure 2(d) shows seven STS spectra recorded at different positions, as marked by the stars with different colors, in Fig. 2(a). Around the Dirac point, the DOS (the slope of the spectra) is linear in energy, which is similar to that of the pristine graphene. Besides the low-energy linear DOS, these spectra show two dips, which are generally of asymmetric strength, flanking the Dirac point. By taking into account the weak 1D electronic potential in the graphene induced by the quasi-periodic ripples [35], it is expected that the electronic chirality will result in highly

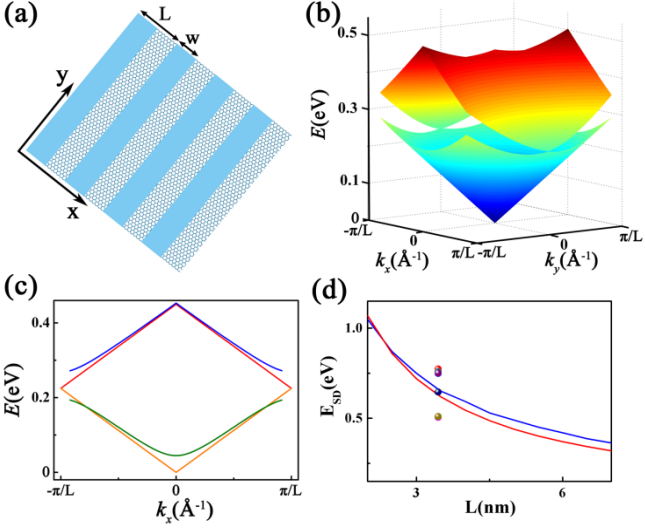

FIG. 3 (color online). (a) The schematic structural model of 1D graphene superlattice formed by a Kronig–Penney type of potential periodic along the x direction with spatial period L and barrier width W. (b) Energy dispersions of charge carriers at the minizone boundaries of a 1D graphene superlattice with $U_{1D}$ = 0.5 eV, L = 10 nm, and W = 5 nm. The gap closes at the minizone boundaries ($k_x$ = $\pm\pi/L$ and $k_y$ = 0) where the superlattice Dirac points are generated. In panel (c) we have a cut of the electronic band structures at a fixed $k_y$. The straight lines correspond to $k_y$ = 0, and the green and blue curves correspond to $k_y$ = 0.012 Å$^{-1}$. The energy gap at the minizone boundaries is zero for $k_y$ = 0. (d) The energy of the superlattice Dirac points away from the Dirac point as a function of the potential period L. The two solid curves are the theoretical dependence of $E_{SD}(L)$ with η = 0.95 (blue) and 0.5 (red), respectively (here η = W/L). The solid circles are the positions of the superlattice Dirac points obtained from the tunneling spectra of Fig. 2(d).

anisotropic behaviors of both the Fermi velocity of the charge carriers and the gap opening at the minizone boundary (MB) formed by the periodic potential [13]. Therefore, the new Dirac points at energies $E_{SD}$ = $\pm\hbar v_F|\mathbf{G}|/2$ can be generated by the 1D potential at the centre of the MB. The two dips in the tunneling spectra are attributed to the positions of the superlattice Dirac points in the DOS. The asymmetric strength of the two dips in the tunneling spectra, as shown in Fig. 2(d), was also observed in the superlattice Dirac points generated by 2D moiré potentials. The asymmetry was mainly attributed to the electron-hole asymmetry originating from next-nearest-neighbor hopping [25]. In strained graphene structures, such large electron-hole asymmetry is expected to observe because of the lattice deformation, which enhances the next-nearest-neighbour hopping [31-33]. Similar asymmetry of the STS spectra is also observed in other corrugated graphene (see Fig. S2 in the supplemental material [42] for STM and STS of another corrugated graphene with similar quasi-periodic ripples).

To further understand the experimental result, we compare our experimental data with the expected theoretical result quantitatively. For simplicity, we assume that the quasi-periodic ripples generate a weak 1D Kronig–Penney type of potential on graphene periodic along the x direction with spatial period L and barrier width W, as shown in Fig. 3(a). Here we assume that L equals to the nominal width of the ripples. The intervalley scattering of such a system may be neglected because of that L is much larger than the nearest-neighbor carbon-carbon distance. Then we can use the model developed in Ref. [13] to show the new Dirac points induced by the 1D potential and derive the period dependence of $E_{SD}$ (see the supplemental material [42] for details of calculation). Fig. 3(b) and (c) show energy dispersions of charge carriers at the minizone boundaries of a 1D graphene superlattice. The group velocity is not renormalized and the energy gap at the MB vanishes when $\mathbf{k}$ is along the direction of the periodic potential, i.e., $k_y$ = 0. Therefore, the 1D graphene superlattice generates four new Dirac points at energies $E_{SD}$ = $\pm\hbar v_F|\mathbf{G}|/2$, which contribute to the two dips in the tunnelling DOS. Our analysis also indicates that the ratio of W/L only influences the position of the superlattice Dirac points slightly, as shown in Fig. 3(d).

The experimental $E_{SD}$ of different ripples of the corrugated graphene shows a wide distribution of values, as shown in Fig. 2(d). We carefully examined all the STS spectra and their measured positions. At a fixed measured position, the value of $E_{SD}$ is almost a constant irrespective of the STM tips and the measured times. However, at different positions, the value of $E_{SD}$ is space-dependent and shows a wide distribution of values. It suggests that the space-dependent of $E_{SD}$ is an intrinsical phenomenon in the corrugated graphene. Similar space-dependent superlattice Dirac points was also observed in each ripple of the corrugated graphene. This excludes the different width of the ripples as a dominating origin of the space-dependent $E_{SD}$. We attribute the observed space-dependent $E_{SD}$ mainly to the space-dependent Fermi velocity, which may originate from the local strain [44] and the strong electron-electron interaction [45,46], in the corrugated graphene according to $E_{SD}$ = $\pm\hbar v_F|\mathbf{G}|/2$. The model developed in Ref. [13] treated the ripples as the quasi-periodic electronic superlattice. The effects of both the strain and the electron-electron interaction were not taken into account in the model. However, recent studies pointed out that the strain and the electron-electron interaction influence the Fermi velocity of graphene dramatically [44-47]. For example, the Fermi velocity is only about $1.0\times10^6$ m/s for graphene without electron-electron interaction or with a very weak electron-electron interaction, whereas, it could reach as high as $3.0\times10^6$ m/s for graphene with strong electron-electron interaction [46]. Actually, the ripples of a corrugated graphene were predicted to lead to a strong electron-electron interaction [18,19,45]. Additionally, it was also predicted theoretically that the corrugated graphene will have a space-dependent Fermi velocity [44].

A best way to confirm the above analysis is to direct measure the space-dependent Fermi velocity in the corrugated graphene. However, this is not easy to realize experimentally. Here we propose a possible solution to explore the space-dependent Fermi velocity in the corrugated graphene. In the pristine graphene, the DOS per unit cell around the Dirac point is given by $\rho(E) \propto |E|/v_F^2$ [3], which indicates that the slope of the DOS reflects the magnitude of the Fermi velocity. Please

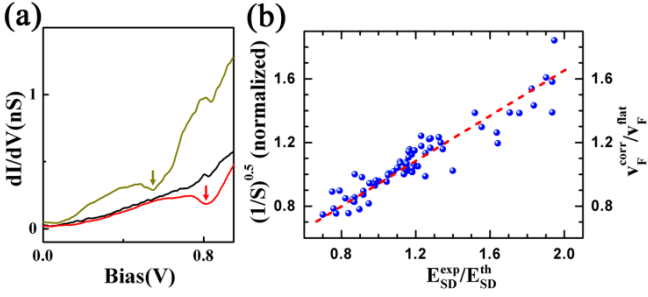

FIG. 4 (color online). (a) Two typical dI/dV-V curves recorded at different positions of the corrugated graphene (the two curves are reproduced from that of Fig. 2(d) with the same colors) along with the STS spectrum of the flat graphene monolayer. Only the spectra in positive bias are shown for clarity. The arrows point to the positions of the superlattice Dirac points observed in the corrugated graphene. (b) The value of $(1/S)^{0.5}$, deduced from the spectra at different positions of different corrugated graphene, as a function of $E_{SD}^{exp}/E_{SD}^{th}$. The right Y-axis of the Fig. 4(b) shows $v_F^{corr}/v_F^{flat}$. Here S is the normalized slope of a STS spectrum and S = 1 corresponds to the slope of the STS spectrum of pristine graphene, $E_{SD}^{exp}$ is the corresponding position of the superlattice Dirac points, $E_{SD}^{th}$ is the theoretical position of the superlattice Dirac points calculated with the width of the ripples L determined experimentally and W/L= 0.5. The red dashed line is the guide to eyes.

note that both theoretical [13] and our experimental results demonstrated that the charge carriers of the graphene superlattice still show a linear increase in the low-energy DOS. Theoretically, the slope of the DOS in graphene under 1D periodic potential is larger than that of the pristine graphene because of the velocity renormalization in the graphene superlattice [13]. It suggests that the slope of the DOS still reflects the magnitude of the average Fermi velocity in the graphene superlattice. In the STS measurements, the slope of the STS spectrum, S, is proportional to $\rho(E)$. On the assumption that the slope of the DOS of the corrugated graphene still reflects the magnitude of the Fermi velocity, i.e., $S \propto 1/v_F^2$ (or $v_F \propto (\frac{1}{S})^{0.5}$), then we can use the STS spectra to derive the local Fermi velocity according to the slope of the spectrum.

If the above assumption is valid, the value of $E_{SD}$ should be proportional to the value of $(1/S)^{0.5}$ according to $E_{SD} = \pm \hbar v_F |\mathbf{G}|/2$. Our experimental result reveals that the value of $E_{SD}$ is closely related to the slope of the STS spectrum, as shown in Fig. 4(a). The two typical curves recorded in the corrugated graphene indicate that the value of $E_{SD}$ is larger at position where the slope of the STS spectrum is smaller. (See Fig. S3 in the supplemental material [42] for method to determine the slope of the STS spectrum. For more experimental evidence, please see Fig. S4 in which we plot the slope of the STS spectra, deduced from the spectra at different positions of different corrugated graphene samples, as a function of the position of the superlattice Dirac points.) Fig. 4(b) shows the value of $(1/S)^{0.5}$ as a function of $E_{SD}^{exp}/E_{SD}^{th}$. The linear dependence of $(1/S)^{0.5}$ and $E_{SD}^{exp}/E_{SD}^{th}$ indicates that the slope of the DOS of the corrugated graphene really reflects the magnitude of the Fermi velocity. The slope of the STS spectra recorded at different positions show a wide distribution of values, which is attributed to the effect of the local strain and the strong electron-electron interaction in the corrugated graphene. The right Y-axis of the Fig. 4(b) shows $v_F^{corr}/v_F^{flat}$. Here $v_F^{corr}$ is the local Fermi velocity in the corrugated graphene and $v_F^{flat}$ is the Fermi velocity in the flat graphene monolayer (these values are obtained on the assumption that $v_F \propto (\frac{1}{S})^{0.5}$). The average Fermi velocity in the corrugated graphene is estimated as $1.1 v_F^{flat}$.

In summary, we address the electronic structures of corrugated graphene monolayer on Rh foil. We show that the quasi-periodic ripples of nanometer wavelength give rise to a weak 1D electronic potential in graphene, which leads to the emergence of the superlattice Dirac points. Our experimental results further demonstrate that the corrugated graphene has a space-dependent Fermi velocity originating from the strain and the electron-electron interaction. These results suggest that the strain and the electron-electron interaction play a vital role in the electronic properties of corrugated graphene and this system could provide an ideal platform to study strongly correlated phases in graphene with desirable properties.

This work was supported by the National Natural Science Foundation of China (Grant No. 11004010, No. 10804010, No. 10974019, No. 21073003, No. 51172029 and No. 91121012), the Fundamental Research Funds for the Central Universities, and the Ministry of Science and Technology of China (Grants No. 2011CB921903, No. 2012CB921404, No 2013CB921701).

§ These authors contributed equally to this paper.
*Email: helin@bnu.edu.cn.